\documentclass[10pt, conference, letterpaper]{IEEEtran} 

\AtBeginDocument{%
  \providecommand\BibTeX{{%
    \normalfont B\kern-0.5em{\scshape i\kern-0.25em b}\kern-0.8em\TeX}}}
    
\usepackage[T1]{fontenc}
\usepackage{graphicx}
\usepackage{float}
\graphicspath{ {images/} }
\usepackage{dsfont}
\usepackage{amsmath}
 
\usepackage[ruled,vlined]{algorithm2e}
\usepackage{amsthm}

\usepackage[font=small]{caption}
\usepackage{subcaption}
\usepackage[export]{adjustbox}
\usepackage{siunitx}
\usepackage{colortbl}
\usepackage{algorithmic}
\usepackage[thinlines]{easytable}
\usepackage{siunitx}
\usepackage[normalem]{ulem}
\usepackage{amsfonts}
\usepackage{multirow,verbatim}
\usepackage[normalem]{ulem}
\usepackage[dvipsnames]{xcolor}
\usepackage{url}
\usepackage{tikz}
\usetikzlibrary{tikzmark}
\usepackage{multirow}
\usepackage{amsthm}
\usepackage[english]{babel}

\usepackage{booktabs,makecell,multirow,hhline,threeparttable} 
\usepackage{xcolor,colortbl,pgf}
\definecolor{Color1}{RGB}{240, 240, 240}
\usepackage{makecell}
\usepackage[top=0.7in, bottom=1.0in, left=0.7in, right=0.7in]{geometry}
\usepackage{hyperref}
\usepackage{orcidlink}

\definecolor{lightgray}{RGB}{211,211,211}
\definecolor{LightCyan}{RGB}{224,255,255}

\SetKwComment{Comment}{$\triangleright$\ }{}

\newcolumntype{b}{>{\columncolor{blue!10}}c}
\newcolumntype{y}{>{\columncolor{yellow!10}}c}
\newcolumntype{d}{>{\columncolor{red!7}}c}

\newtheorem{observation}{Observation}

\SetKwInput{KwInput}{Input}                
\SetKwInput{KwOutput}{Output}              

{}

\usepackage{amsmath,amsfonts,bm}

\def\eqref#1{equation~\ref{#1}}

\def\1{\bm{1}}

\def\vx{{\bm{x}}}
\def\vy{{\bm{y}}}

\def\mI{{\bm{I}}}

\DeclareMathAlphabet{\mathsfit}{\encodingdefault}{\sfdefault}{m}{sl}
\SetMathAlphabet{\mathsfit}{bold}{\encodingdefault}{\sfdefault}{bx}{n}

\newcommand{\E}{\mathbb{E}}
\newcommand{\Ls}{\mathcal{L}}

\newcommand{\KL}{D_{\mathrm{KL}}}

\begin{document}

\title{Waves of Imagination: Unconditional Spectrogram Generation using Diffusion Architectures}

\author{
		\IEEEauthorblockN{
        Rahul Vanukuri\IEEEauthorrefmark{1}, Shafi Ullah Khan\IEEEauthorrefmark{1}, Talip Tolga Sarı\IEEEauthorrefmark{2}, 
        Gokhan Secinti\IEEEauthorrefmark{2}, Diego Patiño\orcidlink{0000-0003-4808-8411}\IEEEauthorrefmark{1},  and
        Debashri Roy\IEEEauthorrefmark{1}
        }

\IEEEauthorblockA{
        \small
\IEEEauthorrefmark{1}The University of Texas at Arlington, USA, \IEEEauthorrefmark{2} Istanbul Technical University, Turkey\\
Emails: \footnotesize{\texttt{rxv6501@mavs.uta.edu;\{sarita,secinti\}@itu.edu.tr;\{shafiullah.khan,diego.patino,debashri.roy\}@uta.edu}}
\vspace{-15pt}
}
}

\maketitle


\begin{abstract}

The growing demand for effective spectrum management and interference mitigation in shared bands, such as the Citizens Broadband Radio Service (CBRS), requires robust radar detection algorithms to protect the military transmission from interference due to commercial wireless transmission. These algorithms, in turn, depend on large, diverse, and carefully labeled spectrogram datasets. However, collecting and annotating real-world radio frequency (RF) spectrogram data remains a significant challenge, as radar signals are rare, and their occurrences are infrequent. This challenge makes the creation of balanced datasets difficult, limiting the performance and generalizability of AI models in this domain.
To address this critical issue, we propose a diffusion-based generative model for synthesizing realistic and diverse spectrograms of five distinct categories that integrate LTE, 5G, and radar signals within the CBRS band. We conduct a structural and statistical fidelity analysis of the generated spectrograms using widely accepted evaluation metrics Structural Similarity Index Measure (SSIM) and Peak Signal-to-Noise Ratio (PSNR), to quantify their divergence from the training data.
Furthermore, we demonstrate that pre-training on the generated spectrograms significantly improves training efficiency on a real-world radar detection task by enabling $51.5\%$ faster convergence.

\end{abstract}


\section{Introduction}
\label{sec:intro}

\noindent{\bf Need for Radio Frequency (RF) Data.} Reliable and timely detection of signals of interest (SoI) is a critical capability in modern wireless systems, underpinning numerous civilian and military applications ranging from dynamic spectrum access and electronic warfare to interference management and situational awareness. Such detection systems increasingly leverage data-driven  machine--learning (ML) methods. Despite this, their effectiveness strongly depends on the availability of sufficiently large and accurately labeled datasets \cite{huang2025raddet}. Unfortunately,  signals characterized by rare, intermittent, or evasive occurrences such as radar emissions operating within the  range of $3.55$--$3.7$ GHz Citizens Broadband Radio Service (CBRS) band, assembling a well-balanced and representative training dataset for field experiments alone becomes impractical \cite{wang2025generative}. To address this critical challenge, synthetic data generation emerges as an essential alternative, providing the necessary volume and variety of training samples \cite{armySBIR}.


\noindent{\bf Representation of RF Data.} Spectrograms efficiently represent joint time–frequency characteristics, offering adequate temporal and spectral resolution together with improved robustness and interpretability over raw In-phase and Quadrature (IQ) representations \cite{9488793}. Moreover, they can be computed in real time with the Short-Time Fourier Transform (STFT) \cite{yan2020comparison}. By capturing key signal dynamics in a 2D format, spectrograms enable mature vision models to be applied to tasks such as detection~\cite{DeepRadar}, localization~\cite{Waldo}, and anomaly detection~\cite{tandiya2018deep}.

\begin{figure}
    \centering
    \includegraphics[width=1\linewidth]{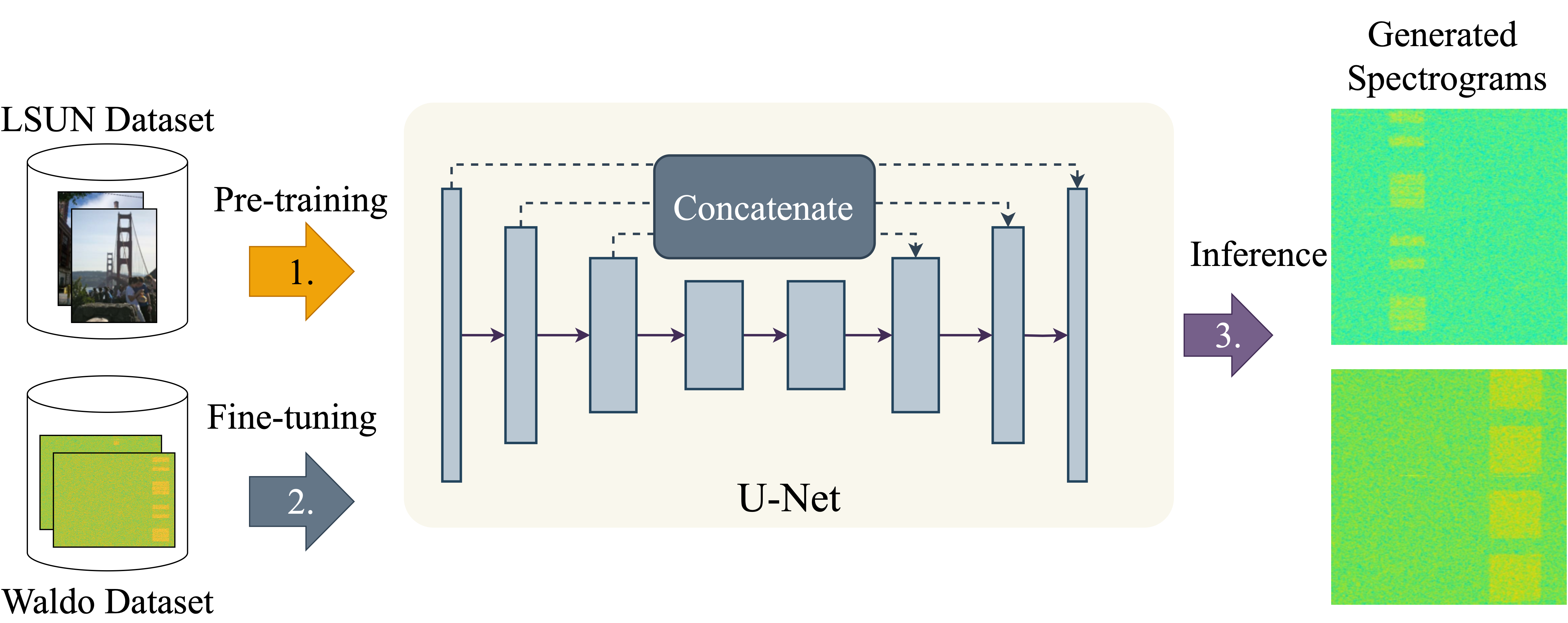}
    \caption{Synthetic spectrogram generation using U-Net~\cite{ronneberger2015u} based diffusion architectures. The LSUN~\cite{yu2015lsun} and Waldo~\cite{Waldo} datasets are publicly available.}
    \vspace{-20pt}
    \label{fig:system}
\end{figure}

\noindent{\bf Challenges of RF Data Collection.} Nevertheless, curating large-scale spectrogram databases that faithfully reflect real-world operating conditions remains an arduous undertaking. A comprehensive corpus must span hundreds of hours of raw captures across disparate geographies, antenna front-ends, propagation conditions, and regulatory constraints while still providing dense annotations that pinpoint the sub-millisecond radar bursts hidden amid much stronger commercial traffic.  {\color{black} It is also possible to create these datasets synthetically, however this requires  expert knowledge to model underlying RF channels accurately to make these datasets realistic~\cite{soltani2019real}. Moreover} manual labeling is labor-intensive because each pixel row corresponds to a narrow frequency slice whose occupancy varies over time; distinguishing genuine radar pulses from out-of-band leakage, harmonics, or impulsive noise often requires expert judgment and cross-referencing with auxiliary sensors~\cite{al2020exposing}. Furthermore, strict spectrum-sharing regulations limit on-air experimentation, and mission-critical incumbents restrict the release of sensitive waveform examples, making public datasets such as Waldo or RadDet orders of magnitude smaller than their computer-vision counterparts~\cite{huang2025raddet, Waldo}. These challenges make it difficult to train deep learning models, which often need tens of thousands of labeled examples to perform well across different devices, environments, and hardware imperfections. To overcome this, we turn to high-fidelity synthetic data generators that can fill the gap by producing realistic and diverse training samples. Specifically, for radar detection in the CBRS band, the generator must be carefully designed to reflect real-world conditions, capturing details such as pulse width, inter pulse interval, and interference patterns based on domain knowledge.

{\color{black}\noindent{\bf Generative Artificial Intelligence (AI) for RF Data Generation.} Generating realistic and diverse spectrogram data, especially for rare-event signals like radar pulses, remains a significant challenge. Several methods for synthetic data generation exist, including analytic channel modeling and generative adversarial networks (GANs). Despite their demonstrated efficacy, each method presents notable limitations. For instance, analytic models often oversimplify the physical world, failing to capture the nuanced, high-dimensional structure of real-world RF signals and their superpositions. Classical GANs, while effective in some image-based domains, frequently suffer from unstable training, hyperparameter sensitivity, and mode collapse, a poor indicator of their utility in modeling rare event distributions~\cite{cao2024survey}.  
Consequently, there is a critical need for a principled, data-driven synthesis framework capable of learning the joint distribution of complex RF environments, supporting the controllable generation of multi-modal, temporally, and spectrally rich signals. Such a framework should (i) produce realistic mixtures of emissions that reflect operational diversity, (ii) capture fine-grained propagation and front-end effects, and (iii) provide quantitative metrics for assessing both perceptual and statistical fidelity, with a demonstrated impact on downstream ML model performance. To address these challenges, we introduce a framework based on denoising diffusion probabilistic model (DDPM) that can generate realistic spectrograms composed of complex signal, including radar, 5G, and LTE pulses. This approach allows us to create diverse high-fidelity training data that closely reflect real-world spectrum environments.

\noindent{\bf Paper Contributions.} In this paper, we propose a diffusion-based generative model tailored to synthesizing realistic spectrogram datasets containing radar, LTE, and 5G signals within the CBRS band. As shown in Fig.~\ref{fig:system}, we employ widely used pre-trained U-Net~\cite{ronneberger2015u} based architecture for diffusion process and generate new spectrogram samples by fine-tuning on a publicly available dataset containing spectrograms of RF signals~\cite{Waldo}. Our novel contributions can be summarized as follows:


\begin{itemize}
    \item We present the first-of-its-kind diffusion-based generative model specifically designed for creating realistic RF spectrograms containing LTE, 5G, and radar signal combinations in shared-spectrum environments.
    \item We conduct a comprehensive evaluation of generated spectrograms using Structural Similarity Index Measure (SSIM) and Peak Signal-to-Noise Ratio (PSNR) to validate fidelity with respect to real-world data.
    \item We demonstrate the practical utility of our synthetic data through a case study showing that machine-learning models trained with generated spectrograms achieve significantly faster convergence compared to models trained solely on limited real-world datasets.
\end{itemize}

\vspace*{-5pt}
\section{Related Work}
\label{sec:related}
Various generative AI frameworks have been proposed to overcome data scarcity in RF sensing and signal detection applications~\cite{chi2024rf}. \cite{wang2025advancing} presents conditional GANs and latent diffusion models to synthesize RF data, including CSI, RFID, and FMCW radar, for tasks such as human activity recognition and 3D pose completion. \cite{chen2023rf} extends this approach by integrating a physics-based ray tracing simulator with cross-modal diffusion models conditioned on vision and text prompts to generate diverse 3D scenes for mmWave radar data. 
\cite{zhang2024generative} contribute SpectrumNet, an open-source dataset enabling generative modeling of multiband 3D radio maps that account for terrain, climate, and frequency diversity.

To support coverage analysis, \cite{sarkar2024recugan} proposes RecuGAN, an information-maximizing GAN for unsupervised generation of diverse radio-frequency coverage maps, while \cite{gibbons2024generative} explores ACGANs and diffusion models for creating synthetic spectrograms that improve bioacoustic classification performance in noisy environments. For radar imaging applications, \cite{xiong2023gpr} introduces GPR-GAN, a GAN-based framework for producing realistic B-scan images in ground-penetrating radar data, demonstrating improved defect detection. Similarly, \cite{hu2022towards} leverages a conditional GAN to reconstruct unconstrained vocabulary audio from vibrations captured by mmWave FMCW radar, converting these vibrations to mel-spectrograms for accurate audio recovery even through barriers.

{\color{black}\noindent{\bf Innovation Opportunity.} However, the state-of-the-art lacks methods that both synthesize high-fidelity spectrograms capturing the transient and multi-standard characteristics of radar, LTE, and 5G signals in shared-spectrum bands and validate their utility on practical detection tasks. In particular, no prior work has explored diffusion probabilistic models specialized for generating spectrograms under CBRS regulatory constraints. Consequently, there remains an unmet need for a generative framework that can produce diverse, physically plausible RF spectrograms and demonstrate their effectiveness in a real-world radar-detection use case.}

\vspace*{-5pt}

\section{System Model \& Proposed Framework}
\label{sec:method}

In this paper, our primary goal is to generate realistic spectrograms containing one or more presence of LTE, 5G, or pulse radar signals. A key aspect of this work is the simulation of scenarios with both isolated and overlapping occurrences of these signals, accurately reflecting the complexity of real-world electromagnetic environments. We use two widely accepted metrics, PSNR and SSIM, to assess the structural and statistical fidelity of the generated spectrograms. Additionally, we conduct a real-world radar detection experiment to evaluate the practical applicability of the generated data. In this experiment, we pre-train a classification model on the generated data to detect five types of radar signals. Later, we fine-tune the classification model using real data and compare the classification accuracy with and without fine-tuning.

\subsection{ Diffusion-based Spectrogram Generation}
In our approach, we adopt an unconditional diffusion framework to generate spectrogram samples as RGB images of size $256 \times 256 \times 3$. The functioning principle of DDPM methods for generating data is two-fold, i.e, the forward and backward processes. First, the forward process incrementally perturbs a clean signal $\vx_0$ over $T$ timesteps, yielding increasingly noisy versions $\vx_1, \ldots, \vx_T$ by adding Gaussian noise with a schedule that controls the variance at each step. The forward process produces $\vx_T$, distributed as standard Normal (Gaussian) distribution, such that $\vx_T \sim \mathcal{N}(\mu, \sigma^2)$. This forward step is fixed since there are no model parameters to train. Second, DDPM (Denoising Diffusion Probabilistic Model) models utilize a trainable machine-learning model to gradually denoise $\vx_T$, structured as a Markov process~\cite{ho2020denoising}, until the original signal is recovered. At inference time, DDPM models only use the denoising part—the backward process to generate new data from a randomly sampled $\vx_T$. Formally, the pipeline is defined as
\begin{equation}
q(\vx_t \mid \vx_0) = \mathcal{N}(\vx_t; \sqrt{\bar{\alpha}_t} \, \vx_0, (1 - \bar{\alpha}_t) \mI),
\label{eq:forward_process}
\end{equation}
where $\bar{\alpha}_t$ denotes the cumulative product of the noise schedule $\alpha_t = 1 - \beta_t$, and $\beta_t$ is a monotonically increasing sequence controlling the aggressiveness of noise injection. This design ensures a smooth transition from the clean data $\vx_0$ to pure noise $\vx_T$ over $T$ steps, thereby setting up a robust foundation for learning the backward mapping.

Each step in this chain is conditionally independent given the previous state, thus establishing a tractable Markov structure. The reverse process, parameterized by a neural network $\boldsymbol{\epsilon}_\theta(\vx_t, t)$, learns to predict and remove the noise component at each timestep, thereby reconstructing the original signal. We train the model to minimize the mean squared error between the true and predicted noise vectors, formalized as the ``simple" loss
\begin{equation}
\mathcal{L}_{\textrm{simple}} = \E_{\vx_0, \epsilon, t} \left[ \left\| \boldsymbol{\epsilon} - \boldsymbol{\epsilon}_\theta(\vx_t, t) \right\|^2 \right],
\label{eq:simple_loss}
\end{equation}
where $\epsilon$ represents sampled Gaussian noise. Empirical evidence has demonstrated that optimizing $\mathcal{L}_{\text{simple}}$ yields samples that are not only visually plausible but also well-aligned with the statistics of the target data distribution.

Next, the reverse denoising process itself is modeled as a conditional Gaussian at each step, with learned mean and variance outputs,
\begin{equation}
p_\theta(\vx_{t-1} \mid \vx_t) = \mathcal{N}(\vx_{t-1}; \mu_\theta(\vx_t, t), \Sigma_\theta(\vx_t, t)),
\label{eq:reverse_param}
\end{equation}
where $\mu_\theta$ and $\Sigma_\theta$ are neural network based projections, ensuring both flexibility and expressivity in modeling the stochastic reverse dynamics.

While the true data likelihood $\log p_\theta(\vx_0)$ is intractable, we maximize a variational lower bound (ELBO), which can be written as
\begin{equation}
\log p_\theta(\vx_0) \geq \E_q \left[ \log \frac{p_\theta(\vx_{0:T})}{q(\vx_{1:T} \mid \vx_0)} \right].
\label{eq:elbo}
\end{equation}
This bound is decomposed into a sum of Kullback–Leibler (KL) divergence terms and a reconstruction log-likelihood, explicitly reflecting the quality of the learned generative process. The overall training objective, often called the variational lower bound (VLB), is thus
\begin{equation}
\small
\begin{aligned}
\Ls_{\text{VLB}} = \E_q \Big[
& \sum_{t=2}^T \KL \left( q(\vx_{t-1} \mid \vx_t, \vx_0) \,\|\, p_\theta(\vx_{t-1} \mid \vx_t) \right) \\
& + \KL \left( q(\vx_T \mid \vx_0) \,\|\, p(\vx_T) \right) - \log p_\theta(\vx_0 \mid \vx_1)
\Big]
\end{aligned}
\label{eq:VLB}
\end{equation}
where each term quantifies the discrepancy between the learned and true reverse transitions. The equation is the combination of terms of previous timestep with its current timestep.

\vspace*{-5pt}

\section{Experiments}
\label{sec:experiments}

\subsection{Spectrogram Generation and Radar Detection Pre-training}

\noindent{\bf Model Architecture.} The general architecture of our data generation strategy follows DDPM. We use a U-Net with residual connections and self-attention modules as a backbone network. We use this backbone for its empirical effectiveness in modeling structured, multi-scale features~\cite{nichol2021improved}. Moreover, we employ two residual blocks at each of the U-Net's resolutions. Feature maps at resolutions (512, 256, 128) $32 \times 32$ and $16 \times 16$ (and optionally $8 \times 8$) are further enriched by multi-head self-attention, enabling the model to capture both local and global dependencies. We introduce sinusoidal timestep embeddings at every residual block, ensuring the network is temporally conditioned and sensitive to the noise level at each denoising step. The encoder progressively reduces the spatial resolution while increasing the depth of feature maps, funneling information through a bottleneck where global context is aggregated. The decoder symmetrically up-samples the feature maps, restoring spatial details via skip connections and final projection to the output space. For models with learned log-variance, the final layer predicts both the denoised image and associated uncertainty, yielding a total of six channels. However, the model only uses the learned variance outputting the three channels as output, which refers to the RGB values of the image. We use a combination of convolutional and ResNet blocks to orchestrate our diffusion architecture~\cite{dhariwal2021diffusion}. We show the representation of our backbone network in Fig.~\ref{fig:Diffusion}.


\begin{figure}[t]
    \centering
    \includegraphics[width=0.8\linewidth,clip,trim=0 0 0 0]{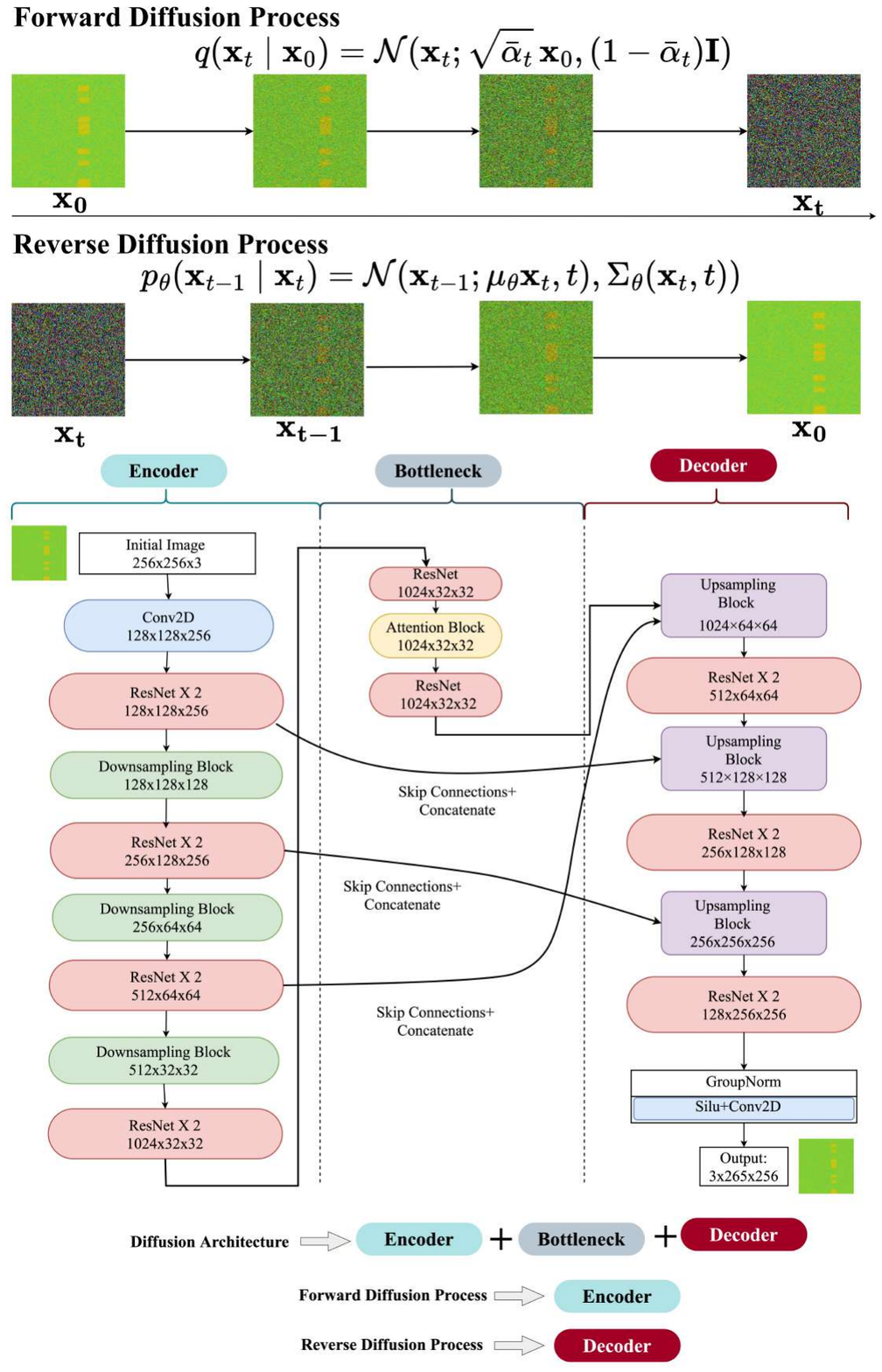}
    \caption{Diffusion model for generating spectrograms. A U-Net-based network~\cite{ronneberger2015u} reverses the noise process to synthesize realistic radar, 5G, and LTE spectrograms.}
    \label{fig:Diffusion}
\end{figure}

\noindent{\bf Pre-training.} We use a pre-trained diffusion model, trained on the Large Science Understanding dataset (LSUN)~\cite{yu2015lsun}. This dataset contains images of $10$ scene categories, including dining room, bedroom, outdoor, church, and others. Each training sample in each category contains between $120,000$ to $3,000,000$ images. The validation data includes $300$ images, and the test data has $1000$ images for each category. The diffusion model, pre-trained on the LSUN dataset, provides checkpoints containing the trained weights.




\subsection{Fine-tuning and Spectrogram Generation Process}
\label{sec:fine-tune}
\noindent{\bf Dataset for Fine-tuning.} We finetune our diffusion-based spectrogram generation model using the publicly available Waldo dataset~\cite{Waldo}. The Waldo dataset comprises $1,229$ spectrogram samples capturing signals within the CBRS band. Each dataset sample covers $100$MHz bandwidth over a $20$ms capture interval. 

\noindent{\bf Fine-tuning Process.} We train the model using data-parallel stochastic gradient descent across multiple GPUs and leverage mixed-precision arithmetic for improved efficiency. We adopt a linear variance schedule with $T=1000$ timesteps and trained for 100 epochs, using the AdamW optimizer with an initial learning rate of $10^{-4}$ and cosine annealing decay. Because the Waldo dataset contains input spectrograms sized at $676\times535\times3$, we resize them to $256 \times 256 \times 3$ to match the model’s input requirements. During training, we maintain exponential moving average weights to stabilize validation performance. We preserve model checkpoints with the lowest validation loss and use them to generate synthetic samples in subsequent experiments.

In summary, we leverage the expressive capacity of diffusion models grounded in rigorous probabilistic inference and equipped with an architecture tailored to the spectral complexity of RF data to establish a robust, scalable foundation for synthesizing high-fidelity, statistically diverse datasets. We select all technical details, including equations for the forward process (Eq.~\ref{eq:forward_process}), reverse process (Eq.~\ref{eq:reverse_param}), and loss formulations (Eq.~\ref{eq:simple_loss} and Eq.~\ref{eq:VLB}), to ensure both theoretical soundness and empirical effectiveness, bridging the gap between simulation and real-world ML-based solution deployments.

\noindent{\bf Spectrogram Generation.} After fine-tuning, we generate $1,223$ spectrograms, categorized into five distinct classes present in the Waldo dataset. The classes we use in our experiments are (a) pure noise, (b) 5G only signals, (c) LTE only signals, (d) composite signals containing 5G with radar, and (d) composite signals containing LTE with radar. A few samples from the generated spectrograms of all the distinct classes (other than the noise) are shown in Fig. \ref{fig:grouped_spectrograms}. Generating each radar sample takes approximately $30$ seconds in an NVIDIA A$4500$ $20$Gb GPU-enabled device.

\begin{figure}[hbt]
\centering
\begin{minipage}{0.9\linewidth}
    \centering
    \begin{subfigure}[]{0.4\linewidth}
        \includegraphics[width=\linewidth]{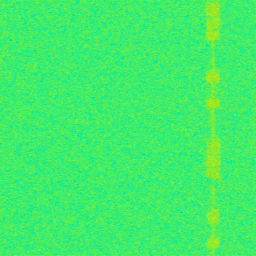}
        \caption{Only LTE}
    \end{subfigure}
    \hfill
    \begin{subfigure}[]{0.4\linewidth}
        \includegraphics[width=\linewidth]{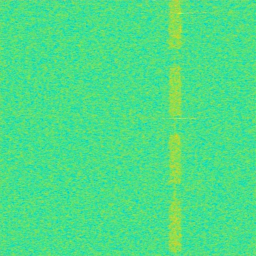}
        \caption{LTE + Radar}
    \end{subfigure}

    \vspace{0.5em}  

    \begin{subfigure}[]{0.4\linewidth}
        \includegraphics[width=\linewidth]{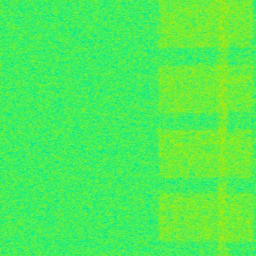}
        \caption{Only 5G}
    \end{subfigure}
    \hfill
    \begin{subfigure}[]{0.4\linewidth}
        \includegraphics[width=\linewidth]{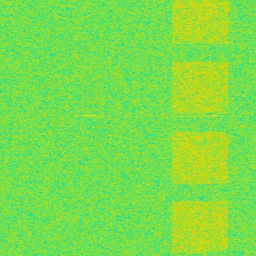}
        \caption{5G + Radar}
    \end{subfigure}
    \caption{Generated spectrogram samples showing the four different categories: LTE, 5G, and Radar combinations.}
    \label{fig:grouped_spectrograms}
    \vspace{-10pt}
\end{minipage}%
\end{figure}

\subsection{Evaluation Metrics}
To quantitatively assess the structural and statistical fidelity of the generated spectrograms against fine-tuning dataset, we use two widely accepted metrics: PSNR and SSIM. PSNR quantifies the pixel-level error between generated and real spectrograms, reflecting absolute signal reconstruction accuracy, whereas SSIM evaluates perceived similarity by capturing visual structural information such as luminance, contrast, and texture similarity.  

\begin{figure}[hbt]
    \centering
    \includegraphics[width=0.48\textwidth]{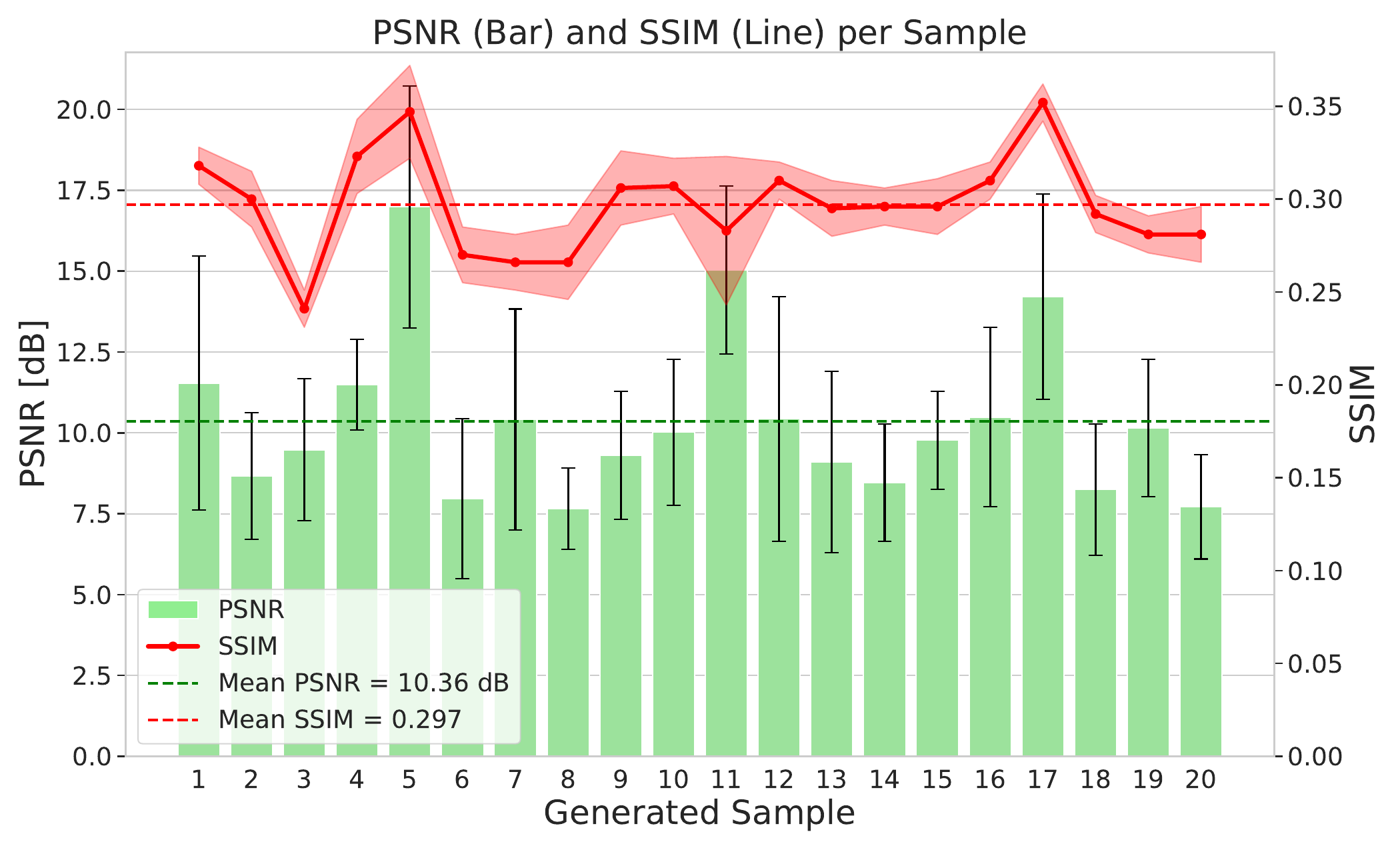}
    \caption{Comparison of 20 randomly generated spectrograms against all the spectrograms of Waldo dataset, presented with 95\% confidence intervals. The PSNR of 10.36\,dB (range 0\,dB (least similar) to $\geq$30\,dB (more similar)) and SSIM of mean 0.29 (range 0 (least similar) to 1 (most similar))  show significant difference in the generated spectrograms from the spectrograms of the Waldo dataset~\cite{Waldo}.}
    \vspace{-10pt}
    \label{fig:spectrogram_diffusionssim}
\end{figure}

\subsection{Evaluation of Generated Spectrograms}

Out of the $1,223$ generated spectrograms, $139$ samples correspond to pure noise, indicating that our framework produces spectrograms containing LTE, 5G, or radar signals in approximately $89\%$ of the cases. Furthermore, $230$ spectrograms exhibit overlapping signal types, which accounts for $19\%$ of the total dataset. To quantitatively assess the similarity between the generated spectrograms and the fine-tuning dataset, we employ the PSNR and SSIM metrics, as shown in Fig.~\ref{fig:spectrogram_diffusionssim}. 

\begin{observation}
The proposed diffusion-based framework successfully generates spectrograms with LTE, 5G, or radar signals in nearly $89\%$ of the cases, demonstrating its reliability in producing meaningful RF signal representations.
\end{observation}

\begin{figure}[hbt]
    \centering
    \vspace{-10pt}
    \includegraphics[width=0.44\textwidth]{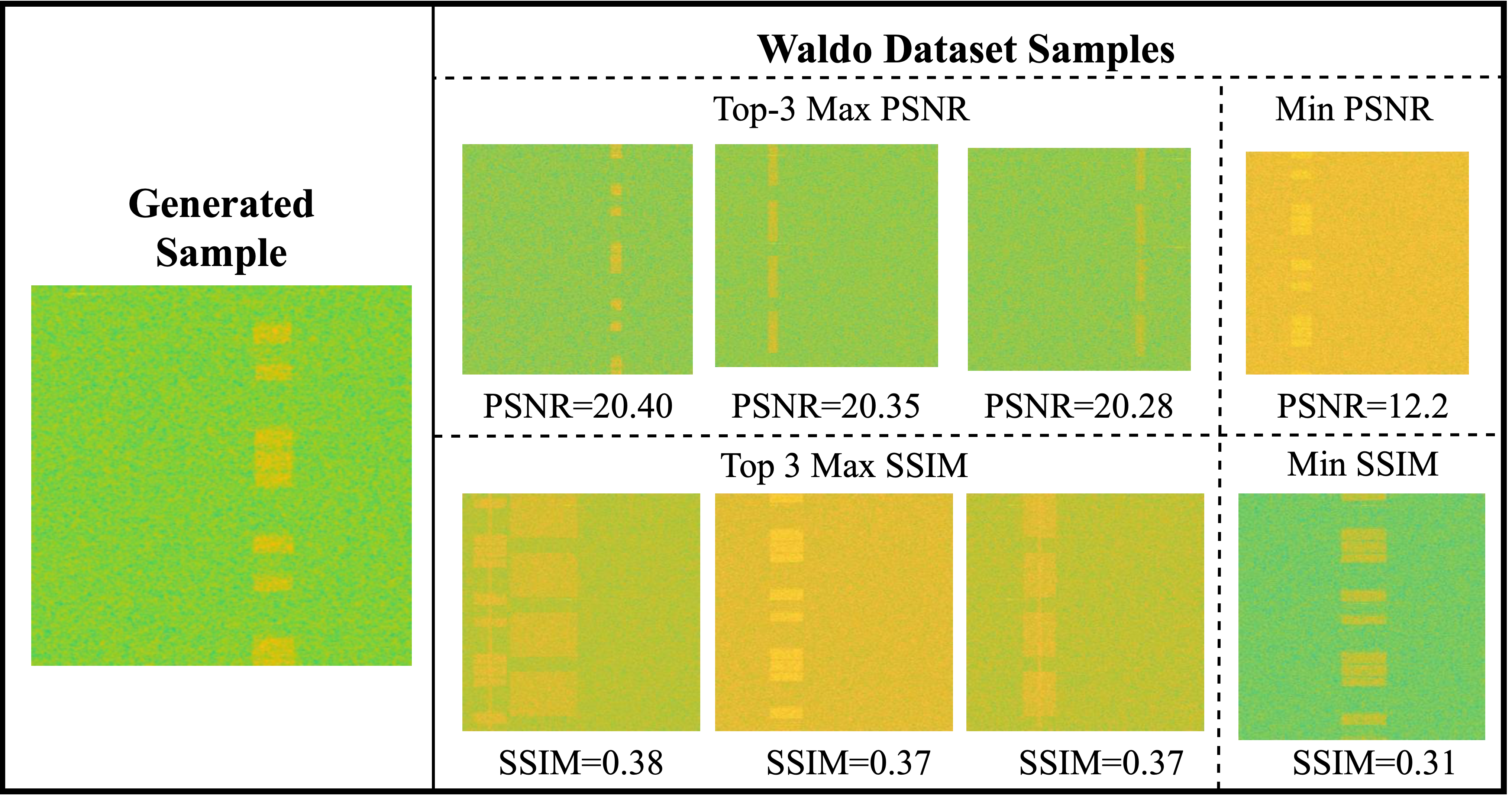}
    \caption{Comparative visual analysis of generated spectrograms versus representative samples from the Waldo dataset. The examples include the top three best and worst matches based on PSNR and SSIM.}
    \label{fig:spectrogram_diffusion}
\end{figure}

To further illustrate the qualitative performance, Fig. \ref{fig:spectrogram_diffusion} presents detailed visual comparisons. We select generated spectrogram samples that achieved the highest and lowest performance according to both PSNR and SSIM metrics. 
PSNR generally ranges from $0$ to $\infty$, with more than $30$\,dB indicating greater similarity between images~\cite{5596999}. SSIM  ranges from $0$ to $1$, where $\geq$$0.9$ indicates greater similarity in terms of structure, content, and luminance~\cite{5596999}. 
Overall, SSIM is given more dominance over the PSNR since we account for both the mean as well as covariances between the pixel values, and the highest similarity is $0.38$, which is the variational sense in terms of structural analysis. The lowest similarity it can produce is up to $31\%$ as shown in Fig.~\ref{fig:spectrogram_diffusion}.

\begin{observation}
We observe the proposed diffusion based framework is able to generate new samples which are significantly different than the training dataset (refer to Fig.~\ref{fig:spectrogram_diffusion}). 
\end{observation}

\section{Usage of Generated Spectrograms}
 In this section, we show how the generated spectrograms can be useful for a real-world radar detection scenario using  transfer learning.
\label{sec:usecase}

\noindent{\bf Scenario: } In our setup, we have an SDR testbed consisting of three USRP B210 devices (Radar TX, 5G TX, and RX) in a controlled laboratory environment. In this setup, the received spectrograms are of one of three types: (a) only radar, (b) only 5G, and (c) radar and 5G. The center frequency is set to the CBRS frequency of $3.65$\,GHz with a constant sampling rate of $50$\,MS/s and a fixed antenna gain of $85$\,dB. The data collection time of each of such samples sums up to $\sim$$5$ minutes for one expert. Hence, this results in a considerable data collection cost for generating large-scale, balanced datasets that can effectively generalize for a simple radar detection task from the received spectrograms. Overall, we collect $450$ training samples, incurring $37.5$ hours of expert data collection time.

\begin{figure}[!hbt]
    \centering
    \vspace{-10pt}
 \includegraphics[width=0.80\linewidth]{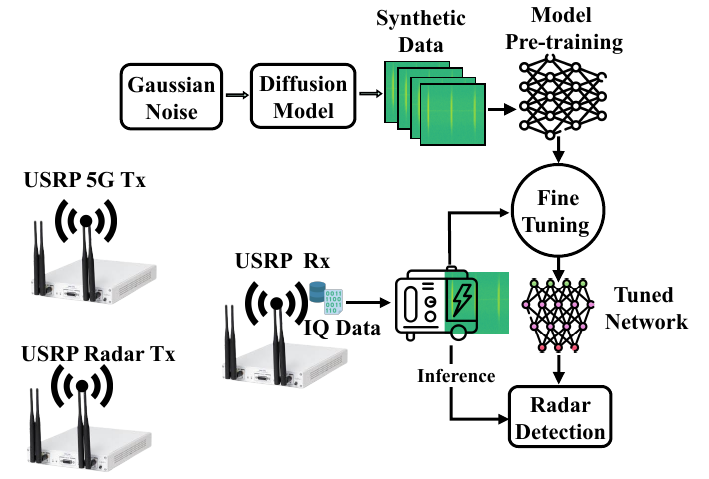}
    \caption{Our experimental setup showing the usage of the generated spectrograms via transfer learning for radar detection use case.}
    \label{fig:tl_setup}
    \vspace{-15pt}
\end{figure}

\noindent{\bf Objective: } Given the limited amount of training data, we aim to train a neural network model to learn the features from the generated spectrograms of section~\ref{sec:experiments} and adapt the domain of these collected signals.

\noindent{\bf Domain Adaption for Transfer Learning: } 
Let $\mathcal{D}_s = \{(\vx_i^s, \vy_i^s)\}_{i=1}^{N_s}$ denote the source domain of generated spectrograms consisting of $N_s$ samples, where $N_s$ = $1223$ from section~\ref{sec:fine-tune}, $\vx_i^s$ and $\vy_i^s$ are the source training samples and labels, respectively. Next, we denote $\mathcal{D}_t = \{(\vx_i^t, \vy_i^t)\}_{i=1}^{N_t}$ as target domain of the real-world collected data with $N_t = 450$ samples and $\vx_i^t$ and $\vy_i^t$ as the target training samples and labels, respectively. The source and target classes are disjoint, i.e., $\mathcal{C}_s \cap \mathcal{C}_t = \emptyset$. Given a pre-trained model with parameters $\theta_s$, our goal is to adapt it to the target domain by optimizing the parameters $\theta_t$ using the following loss function
\begin{equation}
    \mathcal{L}_t(\theta_t) = \frac{1}{N_t} \sum_{i=1}^{N_t} \ell(f(\vx_i^t; \theta_t), \vy_i^t).
\end{equation}
Here, $\ell(\cdot)$ is the classification loss function and $f(\cdot;\theta_t)$ denotes the adapted model. In our case, the pre-trained model $\theta_s$ consists of three convolutional layers and two fully connected layers which classifies among five classes (Noise, 5G, LTE, 5G+Radar, LTE+Radar). The target model shares $\theta_t$ the same model architecture which classifies among three classes (Radar, 5G, 5G+Radar).

\noindent{\bf Result: } 
As shown in Fig.~\ref{fig:accuracy} and Fig.~\ref{fig:loss}, the models pre-trained on synthetic spectrograms show significantly faster convergence compared to models trained from scratch. Specifically, with pre-training, convergence is achieved around epoch $32$, whereas training from scratch requires approximately $66$ epochs to reach stability, representing a $51.5\%$ improvement in convergence speed. These results highlight the utility of our diffusion-generated spectrograms in initializing feature-rich models that generalize effectively even under real-world data scarcity.

\begin{figure}[htbp]
    \centering
\vspace{-10pt}
    \begin{subfigure}[b]{0.40\textwidth}
        \centering
        \includegraphics[width=\textwidth]{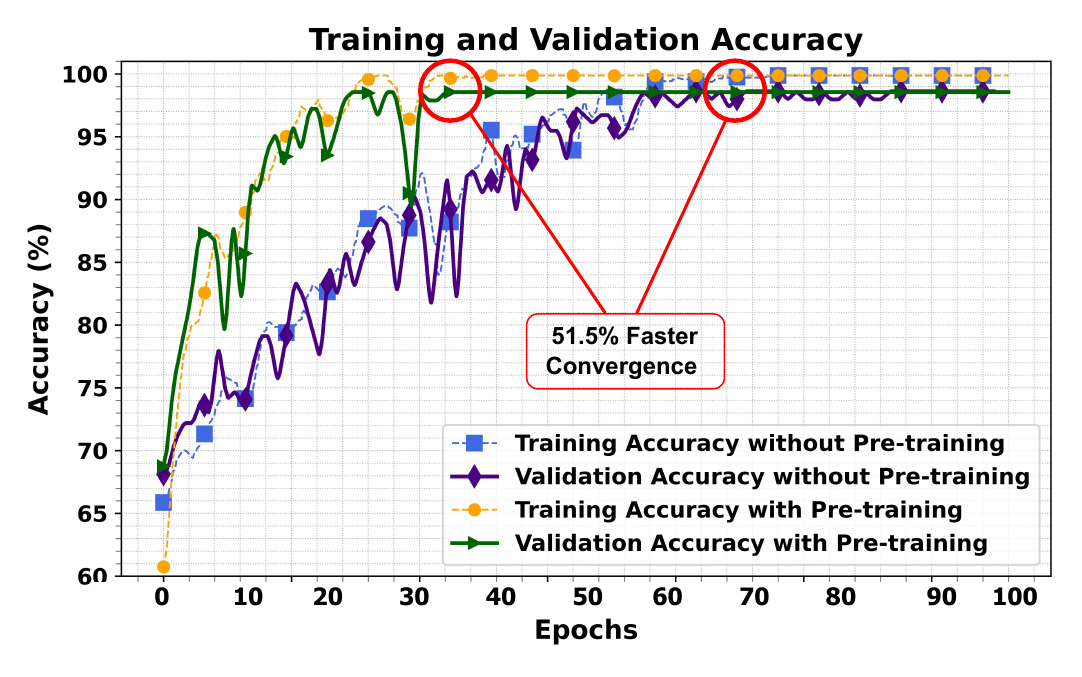}
        \vspace{-20pt}
        \caption{Accuracy curve}
        \label{fig:accuracy}
    \end{subfigure}
    \hfill
    \begin{subfigure}[b]{0.40\textwidth}
        \centering
        \includegraphics[width=\textwidth]{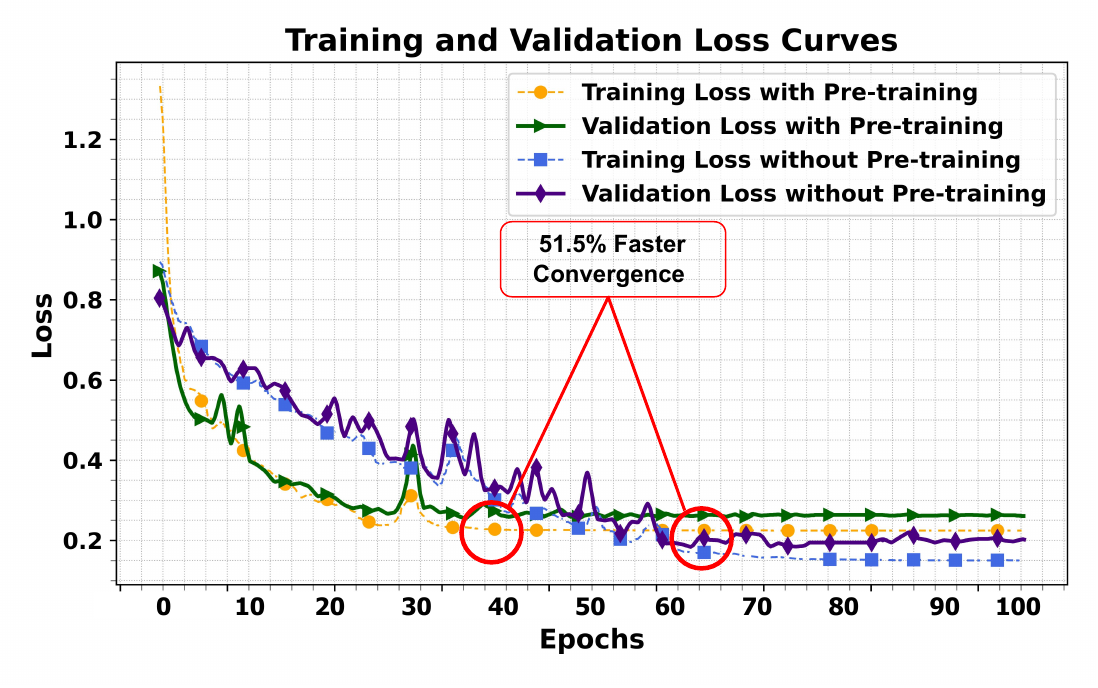}
        \vspace{-20pt}
        \caption{Loss curve}
        \label{fig:loss}
    \end{subfigure}
    \caption{Comparison of training and validation performance w/ and w/o pre-training. Models initialized with diffusion-generated spectrograms converge $51.5\%$ faster. }
    \label{fig:Curves}
     \vspace{-15pt}
\end{figure}

\vspace*{-5pt}
\begin{observation}
We observe the pre-training using the generated spectrograms can help to minimize the training time or new data requirement to attain a specific performance for a new type of radar detection use case (see Fig.~\ref{fig:Curves}).
\end{observation}

\vspace*{-5pt}

\section{Conclusion}
\label{sec:conclusion}
In this work, we address the persistent challenge of creating realistic and diverse RF spectrogram datasets, particularly for rare-event signals such as radar pulses in shared-spectrum environments. By leveraging advanced diffusion probabilistic models, we successfully synthesize spectrograms containing intricate combinations of radar, LTE, and 5G signals.
Experimental validation confirmed that our synthetic spectrograms exhibit high structural fidelity and statistical similarity compared to real-world data, as quantified through SSIM and PSNR metrics. Moreover, practical ML models trained on our synthetic dataset demonstrated accelerated convergence and enhanced detection performance. 
These results highlight our framework's practicality, scalability, and effectiveness, marking a substantial advancement toward realistic, robust RF signal detection. Looking forward, we aim to expand our methodology to more signal types and with user commands to generate spectrograms with signals of their choice.

\vspace*{-10pt}

\section*{Acknowledgment}
\vspace{-0.2cm}
\label{sec:ack}
The authors gracefully acknowledge the sponsorship from US National Science Foundation (CNS 2526490).
\vspace{-0.4cm}


\bibliographystyle{IEEEtran} 
\bibliography{reference}

\end{document}